\documentclass[pra,twocolumn,showpacs,preprintnumbers]{revtex4}
\usepackage{amsmath}
\usepackage{graphicx}
\usepackage{mathrsfs}
\usepackage{amssymb}
\usepackage{amsfonts}
\usepackage{amsmath}
\usepackage[colorlinks,citecolor=blue,linkcolor=red,hyperindex,CJKbookmarks,dvipdfm]{hyperref}
\begin{document}
\title{Resonator-Assisted Quantum Bath Engineering of a Flux Qubit}
\author{Xian-Peng Zhang$^{1}$}
\author{Li-Tuo Shen$^1$}
\author{Zhang-Qi Yin$^2$}
\author{Huai-Zhi Wu$^1$}
\author{Zhen-Biao Yang$^1$}
\email{zbyang@fzu.edu.cn}
\affiliation{1.Department of Physics, Fuzhou University, Fuzhou, 350108, P. R. China}
\affiliation{2.Center for Quantum Information, Institute for Interdisciplinary Information Sciences, Tsinghua
University, Beijing 100084, P. R. China}

\begin{abstract}
We demonstrate quantum bath engineering for preparation of any orbital state with controllable phase factor of a superconducting flux qubit assisted by a microwave coplanar waveguide resonator. We investigate the polarization efficiency of the arbitrary direction rotating on the Bloch sphere, and obtain an effective Rabi frequency by using the convergence condition of Markovian master equation. The processes of polarization can be implemented effectively in a dissipative environment created by resonator photon loss when the spectrum of the microwave resonator matches with the specially tailored Rabi and resonant frequencies of the drive. Our calculations indicate that state-preparation fidelities in excess of 99\% and the required time on the order of magnitude of microsecond are in principle possible for experimentally reasonable sample parameters. Furthermore, our proposal could be applied to other systems with spin-based qubits.
\end{abstract}
\maketitle
One of the most promising achievements from the exploration of the hybrid quantum circuits is harnessing the advantages of the different quantum systems to discover the new qualities that are not acquirable for either independent system \cite{RMP2013-85-553,RMP2013-85-623}. An exemplification is photon-participated initialization of atom, spin and superconducting qubits. Manipulation of genuine quantum systems requires that they should be effectively prepared into a well-defined quantum state, which is not only important for quantum error correcting of quantum information processors \cite{S2011-332-1059,N2012-482-382} but is also of significance for the applications in enhancing quantum memories \cite{PRL2010-105-140501,PRL2010-105-140502}.

In theory, any qubit can be prepared into its minimal energy state, i.e., ground state, when cooling to so low temperature that thermal excitation energy is much less than the energy splitting of the qubit. Consequently, low temperature environment is bound to slow down systems to reach the thermodynamic equilibrium, which retards the operations in quantum information processors \cite{OUP1961}. More effective cooling schemes have been studied extensively in the context of Doppler and Sisyphus cooling \cite{PRL2003-90-133602,NP2008-4-612}, algorithmic cooling \cite{N2011-480-500,PRL2008-100-140501}, cavity cooling \cite{NP2005-1-122,NL2004-428-50,PRL2014-112-050501}, etc. The method of cavity cooling (say, for atomic gases \cite{NL2004-428-50,PRL2009-103-103001}, mechanical objects \cite{NL2006-444-71,NL2006-443-193}, spins \cite{PRA2010-82-041804,PRL2014-112-050501}, etc.) that utilizes the way to dissipate the kinetic energy in open environment created by cavity photon loss in a controlled manner has been investigated. Currently, it was demonstrated that a superconducting transmon qubit may be prepared in any pure state of the Bloch sphere with high fidelity assisted by a microwave cavity \cite{PRL2012-109-183602}. However, the phase factor of the prepared state is uncontrollable.

We present in this paper a scheme for preparation of any orbital state with controllable phase factor of
a superconducting flux qubit including three mesoscopic Josephson junctions arranged in a superconducting loop assisted by a single-mode coplanar waveguide (CPW) resonator. In particular, we investigate the polarization efficiency of the arbitrary rotations on the Bloch sphere and obtain an effective Rabi frequency which depends on its polarization direction by using the convergence condition of Markovian master equation. The processes of polarization can be implemented rapidly enough in the direction where the spectrum of resonator matches with the specially tailored Rabi and resonant frequencies of the drive, which is essential for the state preparation of a superconducting flux qubit by adjusting system parameters. Our calculations indicate that state preparation fidelities in excess of 99\% and the required time on the order of magnitude of microsecond are in principle possible with currently achievable sample parameters, which is significantly shorter than the thermal relation time for the low-temperature superconducting flux qubit \cite{arXiv:1407.1346v1,arXiv:1403.3871v2}. Furthermore, our scheme could be applied to other kinds of superconducting qubits, as well as to other physical systems.
\begin{figure*}[htbp]
\centering
\includegraphics[width=1.8 \columnwidth]{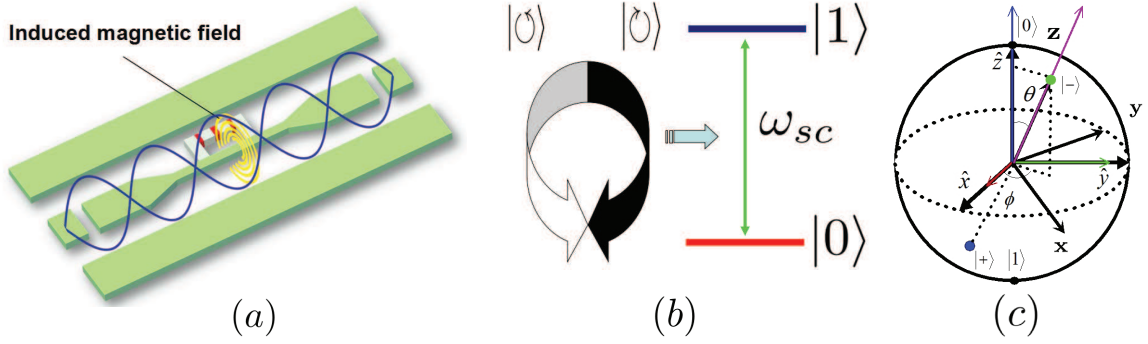}
\caption{(Color online) (a) A superconducting flux qubit is coupled to a CPW resonator via the induced magnetic field. The blue sinusoidal curves describe the microwave drive. (b) The superconducting flux qubit is labeled with the new eigenstates $\left\vert0\right\rangle$ and $\left\vert1\right\rangle$ with the energy splitting $\omega_{sc}$. (c) Bloch sphere diagrams indicate that polarization direction \textbf{z} (pink arrow) defined by an around-z-axis rotation with angle $\phi$ -$R_z(\phi)$ followed by an around-y-axis rotation with angle $\theta$ -$R_y(\theta)$, is determined by the rate of the detuning of the drive (blue arrow), the real part (red arrow) and imaginary part (green arrow) of the Rabi frequency (as illustrated by equation (\ref{e20})).} \label{f1}
\end{figure*}

We here consider a superconducting flux qubit comprising three mesoscopic Josephson junctions in a loop (depicted in FIG. \ref{f1}(a))
threaded by an induced magnetic field \cite{S1999-285-1036}. The flux qubit couples to a CPW resonator via the induced magnetic field \cite{NP2010-6-772,PRL2010-105-023601}. As shown in FIG. \ref{f1}(b), two computational basis states of the flux qubit carry opposite macroscopic persistent currents. The flux qubit can be described by the effective Hamiltonian $H_{SC} =-\left(B_z\tilde{\sigma}_z+B_x\tilde{\sigma}_x\right)/2$, where $\tilde{\sigma}_{z,x}$ are the Pauli matrices, $B_x$ is the level repulsion, $ B_z$ is the DC energy bias, and the rewritten qubit levels $\left\vert 0 \right\rangle $ and $\left\vert 1 \right\rangle $ have energies $\mp\frac{1}{2}\omega_{sc}$ ($\omega_{sc}=\sqrt{B_x^2+B_z^2}$) respectively ($\hbar=1$ is used throughout this paper). In the presence of a microwave drive, Rabi oscillations between energy levels $\left\vert 0 \right\rangle $ and $\left\vert 1 \right\rangle $ are induced near resonance.  The total Hamiltonian of joint system is taken as
$H=H_{0}+H_{d}+H_{r}$ with
\begin{eqnarray}
  H_{0}    &=& \omega_c a^{\dagger}a+\frac{\omega_{sc}}{2}\sigma_z,\\
  H_{d}    &=& \Omega \sigma_-e^{i\varpi_L t}+ \tilde{\Omega}\sigma_-e^{-i\varpi_L t}+H.c.,\\
  H_{r}    &=& g\left(a+a^{\dagger}\right)\sigma_x,
\end{eqnarray}
where $a\left(a^{\dagger}\right)$ are the annihilation (creation) operators of the CPW resonator with frequency $\omega_c$ and linewidth $\kappa$, $\Omega$ and $\tilde{\Omega}$ are the rotating and counter-rotating Rabi frequencies of the drive with frequency $\varpi_L$, and $g$ is the light-qubit coupling constant. Here we use the Pauli operators $\sigma_\imath(\imath=x,y,z,\pm)$ for the flux qubit (with the qubit ground and excited states, $\left\vert 0 \right\rangle $ and $\left\vert 1 \right\rangle $), $\sigma_{\pm}$ are the raising (lowering) operators, and $\sigma_{x,y,z}$ are the $x,y,z$-Pauli operators.

In the interaction picture with the rotating Hamiltonian $H_1= H_0-\delta\omega a^{\dagger}a-\delta\varpi \sigma_z/2$, the Hamiltonian of the composite system within the standard rotating wave approximation (RWA) is
\begin{eqnarray}\label{e4}
  \tilde{H}_{1} &=& ga^{\dagger}\sigma_-+ga\sigma_++\delta\omega a^{\dagger}a+\frac{\delta\varpi}{2} \sigma_z \notag\\
      &+&\texttt{Re}(\Omega)\sigma_x+\texttt{Im}(\Omega)\sigma_y,
\end{eqnarray}
with $\delta\varpi=\omega_{sc}-\varpi_L$ and $\delta\omega=\omega_c-\varpi_L$. This RWA is enforced in the parameter regime $\omega_c,\varpi_L,\omega_{sc} \gg g,\kappa,\Omega,\tilde{\Omega}$.

Assume that the flux qubit should be prepared in any arbitrary superposition of ground and excited states on demand:
\begin{equation}\label{e5}
  \left\vert - \right\rangle=\cos(\frac{\theta}{2})\left\vert 0 \right\rangle +e^{i\phi}\sin( \frac{\theta}{2})\left\vert 1 \right\rangle
\end{equation}
with $\theta\in[0 ,\pi]$ and $\phi\in [0,2\pi)$, which is the eigenstate of the Pauli operator component
\begin{equation}
  \sigma_\textbf{z}=-\sin\theta\cos\phi \sigma_x+\sin\theta\sin\phi \sigma_y+\cos\theta \sigma_z
\end{equation}
with eigenvalue $-1$. While the other eigenstate with eigenvalue $+1$ is $\left\vert + \right\rangle=\sin(\frac{\theta}{2})\left\vert 0 \right\rangle -e^{i\phi}\cos( \frac{\theta}{2})\left\vert 1 \right\rangle$. Through weakly coupling to a resonator as well as to a microwave drive, the qubit can be polarized to the $\left\vert + \right\rangle$ or $\left\vert - \right\rangle$ state.

To investigate the polarization efficiency, we introduce a unitary transformation, $\textbf{R}(\theta,\phi)$, for Pauli operators:
\begin{equation}
  \left[
    \begin{array}{c}
      \sigma_{\textbf{x}} \\
      \sigma_{\textbf{y}} \\
      \sigma_{\textbf{z}} \\
    \end{array}
  \right]
   =\left[
     \begin{array}{ccc}
       \cos\theta\cos\phi & -\cos\theta\sin\phi & \sin\theta \\
       \sin\phi & \cos\phi & 0\\
       -\sin\theta\cos\phi & \sin\theta\sin\phi & \cos\theta
     \end{array}
   \right]
  \left[
    \begin{array}{c}
      \sigma_x \\
      \sigma_y \\
      \sigma_z \\
    \end{array}
  \right].
\end{equation}
As illustrated in FIG. \ref{f1}(c), this unitary transformation corresponds to a space rotation of Pauli operation defined by an around-z-axis rotation with angle $\phi$ -$R_z(\phi)$ followed by an around-y-axis rotation with angle $\theta$ -$R_y(\theta)$. From here, the bold subscripts $\textbf{x}, \textbf{y}, \textbf{z}$ indicate the space basics after the rotation. After moving into the interaction frame of $H_2=\bar{\Omega}\sigma_{\textbf{z}}+\delta\omega a^{\dagger}a$, the Hamiltonian (\ref{e4}) transforms to
\begin{eqnarray} \label{e8}
   \tilde{H}_{2}(t)&=& (A_{\textbf{z}}-\bar{\Omega})\sigma_{\textbf{z}}+\tilde{H}_{\bar{\Omega}}(t)
       +\tilde{H}_{\textbf{z}}(t) +\tilde{H}_{-}(t) +\tilde{H}_{+}(t),\nonumber\\
       \\
   \tilde{H}_{\bar{\Omega}}(t) &=& (A_{\textbf{x}}-iA_{\textbf{y}})e^{i2\bar{\Omega}t}\sigma_+^{(\textbf{z})}
       +H.c.,\\
   \tilde{H}_{\textbf{z}}(t) &=& \Theta_{\textbf{z}}e^{i\delta\omega t}ga^{\dagger}\sigma_{\textbf{z}}
       +H.c., \\
   \tilde{H}_{-}(t) &=& \Theta_+e^{i(\delta\omega-2\bar{\Omega})t}ga^{\dagger}\sigma_-^{(\textbf{z})}
       +H.c., \\
   \tilde{H}_{+}(t) &=& \Theta_-e^{i(\delta\omega+2\bar{\Omega})t}ga^{\dagger}\sigma_+^{(\textbf{z})}
       +H.c.,
\end{eqnarray}
with
\begin{eqnarray}
 &&[A_{\textbf{x}},A_{\textbf{y}},A_{\textbf{z}}]^T = \textbf{R}[\texttt{Re}(\Omega),\texttt{Im}(\Omega),\delta\varpi/2]^T,\\
 &&[\Theta_{\textbf{x}},\Theta_{\textbf{y}},\Theta_{\textbf{z}}]^T = \textbf{R}[1/2,-i/2,0]^T,\\
 &&\Theta_\pm =\Theta_{\textbf{x}}\pm i\Theta_{\textbf{y}},
\end{eqnarray}
which are specified in the Appendix, where $\bar{\Omega}$ is the effective Rabi frequency that will be obtained by using the convergence condition of Markovian master equation, and $\sigma^{(\textbf{z})}_\pm =(\sigma_{\textbf{x}} \pm i\sigma_{\textbf{y}})/2$ are the ladder operators in the $\textbf{z}$-basis.

There is no preference in the $\sigma_{\textbf{z}}$ direction for the dynamics of $\tilde{H}_{\textbf{z}}(t)$ and $\tilde{H}_{\bar{\Omega}}(t)$ at the thermal equilibrium, while those of $\tilde{H}_{\pm}(t)$ would drive the flux qubit to $\left\langle \sigma_{\textbf{z}} \right\rangle=\pm 1$ state, respectively \cite{PRL2014-112-050501}. We may set $\Delta=\delta\omega-2\bar{\Omega}$ to be close to zero, so that the absolute value of $\Delta$ is small as compared to those of $\delta\omega,2\bar{\Omega}$. After making the second RWA in the interaction frame of $H_2$, the interaction Hamiltonian reduces to
\begin{equation}\label{e7-16}
  H_I(t)=(A_{\textbf{z}}-\bar{\Omega})\sigma_{\textbf{z}}+ \Theta_+e^{i\Delta t}ga^{\dagger}\sigma_-^{(\textbf{z})}
       +\Theta^*_+e^{-i\Delta t}ga\sigma_+^{(\textbf{z})}.
\end{equation}
The RWA used here is satisfied when the absolute values of $\delta\omega$ and $\bar{\Omega}$ are large compared to the time scale of interest $(\vert\delta\omega\vert,\vert2\bar{\Omega}\vert \gg \kappa,\vert A_{\textbf{x}}\pm iA_{\textbf{y}}\vert,\vert g\Theta_{\textbf{z}}\vert,\vert g\Theta_-\vert)$.

To obtain the Markovian master equation for the driven flux qubit, we assume the bad resonator condition $\kappa \gg g$. The reduced dynamics of the flux qubit in the interaction frame of the dissipator is given to the second order by the time-convolutionless master equation (see Appendix and Ref. \cite{PRL2014-112-050501}):
\begin{equation}\label{e11-e17}
  \dot{\varrho}(t)=\int_0^{\infty} d\tau \texttt{tr}_c[e^{\tau D^{\dagger}_c}(L[H_I(t)])L[H_I(t-\tau)]\varrho(t)\otimes\rho_{eq}],
\end{equation}
where $L$ is the superoperator $L[X]\rho(t)=-i[X,\rho(t)]$, $\varrho(t) = \texttt{tr}_c[\rho(t)]$ is the reduced state of the flux qubit and $\rho _{eq}$ is the equilibrium state of the resonator.

Using the algebraic transformation of the dissipator $D_{c}$ \cite{PRL2014-112-050501}: $e^{tD_{c}^{\dagger}}[\textbf{I}]=\textbf{I}, e^{tD_{c}^{\dagger}}[a]=e^{-\kappa t/2}a, e^{tD_{c}^{\dagger}}[a^{\dagger}]=e^{-\kappa t/2}a^{\dagger}$, the master equation (\ref{e11-e17}) reduces to
\begin{eqnarray}\label{e9-e18}
  \dot{\varrho}(t)&=&\int_0^{\infty}\{e^{-\kappa\tau/2}\texttt{tr}_c[L[\tilde{H}_{-}(t)]
                  L[\tilde{H}_{-}(t-\tau)]\varrho(t)\otimes\rho_{eq}] \notag\\
                  &+&L[(A_{\textbf{z}}-\bar{\Omega}) \sigma_{\textbf{z}}]L[(A_{\textbf{z}}-\bar{\Omega}) \sigma_{\textbf{z}}]\varrho(t)\}d\tau,
\end{eqnarray}
where the cross terms for the 2nd order TCL master equation have been removed with the properties of our resonator equilibrium state:
$\texttt{tr}_c[a\rho_{eq}]=\texttt{tr}_c[a^{\dagger}\rho_{eq}]=0$. We find that the last term of the master equation (\ref{e9-e18}) will not be
convergent unless the constant of component Hamiltonian, $A_{\textbf{z}}-\bar{\Omega}$, becomes zero. Therefore, we obtain the effective
Rabi frequency
\begin{equation}
  \bar{\Omega}=-\texttt{Re}(\Omega)\sin\theta\cos\phi+\texttt{Im}(\Omega)\sin\theta\sin\phi+\frac{\delta\varpi}{2}\cos\theta.
\end{equation}

Considering the RWA condition $\vert2\bar{\Omega}\vert\gg \vert A_{\textbf{x}}\pm iA_{\textbf{y}}\vert$, we obtain the parameter relationships
\begin{equation}\label{e20}
  -\frac{\sin\theta\cos\phi}{\texttt{Re}(\Omega)}\approx\frac{\sin\theta\sin\phi}{\texttt{Im}(\Omega)}\approx\frac{2\cos\theta}{\delta\varpi}.
\end{equation}
The controllable phase factor of the prepared state, determined by the $\sigma_x$ and $\sigma_y$ parts of the Pauli operator component $\sigma_{\textbf{z}}$, is actually manipulated by the phase of Rabi frequency; while the state populations, tailored by the $\sigma_z$ part, ultimately rely on the rate of $\delta\varpi/\vert\Omega\vert$. Therefore, the preparation of arbitrarily specified coherent superposition of the ground and excited states  of a flux qubit can be implemented by adjusting system parameters $[\texttt{Re}(\Omega),\texttt{Im}(\Omega),\delta\varpi/2]$.

The most efficient polarization for the target state with $\left\langle \sigma_{\textbf{z}} \right\rangle=-1$ happens when the effective Rabi frequency is matched to the spectrum of the resonator, ie., $\delta\omega=2\bar{\Omega}$, where the effective polarization rate becomes
\begin{eqnarray}
 \Gamma_{\textbf{z}}&=& \frac{g^2\kappa(1+\cos\theta)^2}{\kappa^2+4\Delta^2},
\end{eqnarray}
and the master equation (\ref{e9-e18}) reduces to a rate equation for the state populations:
\begin{equation}
  \frac{d}{dt}\vec{P}(t) =\Gamma_\textbf{z} \textbf{M}\vec{P}(t),
\end{equation}
with
\begin{equation}
  \textbf{M}=\left[
        \begin{array}{cc}
          -\bar{n} & \bar{n}+1 \\
          \bar{n} & -(\bar{n}+1) \\
        \end{array}
      \right].
\end{equation}
Here $\bar{n}$ is the average photon number at equilibrium, the diagonal matrix elements $P_{m}(t)=\left\langle m \right\vert\varrho(t)\left\vert m\right\rangle(m =\pm 1)$ of the reduced density operator $\varrho(t)$ corresponds to the expectation value of the projection operator $\left\vert m\right\rangle \left\langle m\right\vert$ at the arbitrary time $t$, and $\vec{P}(t)=(P_{-1}(t),P_{1}(t))^T$ is defined.

At the thermodynamic equilibrium, the state of the driven flux qubit satisfies $\partial_t\vec{P}_{J}(\infty) =0$ and can be given by $\rho_{J,eq}=\sum_{m=\pm 1}P_{m}(\infty)\varrho_{m}$, where
\begin{eqnarray}
  P_{-1}(\infty)&=&\frac{1}{e^{-\omega_c/k_BT_c}+1},\\
  P_{1}(\infty)&=&\frac{e^{-\omega_c/k_BT_c}}{e^{-\omega_c/k_BT_c}+1}.
\end{eqnarray}
The expectation value of the Pauli operator component $\sigma_\textbf{z}$ for the equilibrium state is
\begin{eqnarray}
 \left\langle \sigma_\textbf{z} \right\rangle_{eq} &=&\frac{ e^{-\omega_c/k_BT_c}-1}{e^{-\omega_c/k_BT_c}+1}.
\end{eqnarray}
In the ideal case where the resonator is cooled to its ground state ($T_c\rightarrow0$), the probability of the qubit being in state $\left\langle \sigma_{\textbf{z}} \right\rangle=-1$ at equilibrium is given by $P_{-1}\simeq 1$ and the final expectation value of the Pauli operator component $\sigma_\textbf{z}$ is approximately $\left\langle \sigma_\textbf{z} \right\rangle_{eq}\simeq-1$.

\begin{figure*}[htbp]
\centering
\includegraphics[width=2.08 \columnwidth]{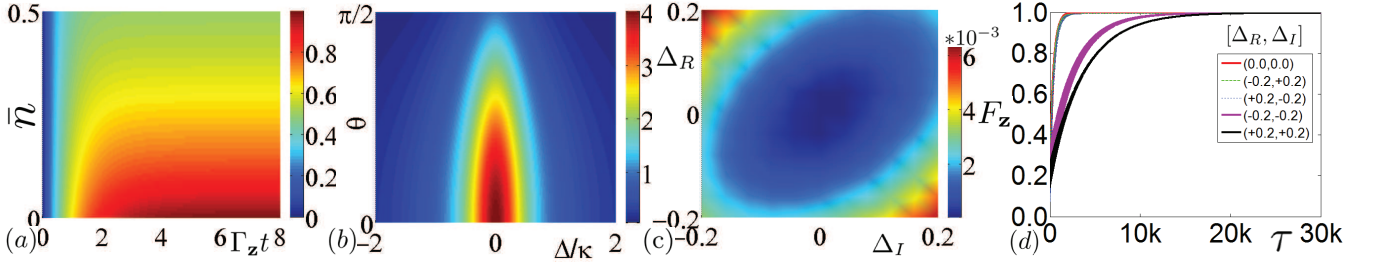}
\caption{(Color online) (a) Normalized expectation value $-\left\langle \sigma_\textbf{z}\right\rangle$ of the flux qubit as a function of the dimensionless parameter $\Gamma_\textbf{z}t$ for various equilibrium temperatures of the resonator ranging from $\bar{n}=0$ to $\bar{n}=0.5$. \label{f2a} (b) Effective dissipation rate in the units of $g^2/\kappa=1$ versus the dimensionless parameters $\Delta/\kappa$ and $\theta$.\label{f2b} (c) The infidelity of the generated state as a function of the dimensionless parameters $\Delta_R=\delta \texttt{Re}(\Omega)/\texttt{Re}(\Omega)$ and $\Delta_I=\delta \texttt{Im}(\Omega)/\texttt{Im}(\Omega)$ at equilibrium, i.e., $IF_\textbf{z}$ vs $\Delta_R$ and $\Delta_I$, for parameters $[\texttt{Re}(\Omega),\texttt{Im}(\Omega),\delta\varpi/2]/2\pi=[100,100,100]/\sqrt{3}$ MHz and $[\theta,\phi]=[\arccos(1/\sqrt{3}),3\pi/4]$. \label{f2c} (d) The evolution of the fidelity of the ground state of $\sigma_{\textbf{z}}=(\sigma_x+\sigma_y+\sigma_z)/\sqrt{3}$ for different deviations of parameters $\Delta_R$ and $\Delta_I$, where the deviation situations of $[\Delta_R,\Delta_I]=[0.0,0.0],[-0.2,+0.2],[+0.2,-0.2]$, which respond to red, green and blue curves, respectively, almost overlap. The dimensionless parameter $\tau=2\bar{\Omega}t$ is defined, while other parameters are the same as (c). Here k means $10^3$. \label{f2d}}
\end{figure*}

Assume that the flux qubit is taken to be maximally mixed in the basis $\{P_{m}(0)=1/2$, for $m=\pm1\}$. The simulated expectation value of $\left\langle \sigma_\textbf{z}(t) \right\rangle$ for the temperature of bath ranging from $\bar{n}=0$ to $\bar{n}=0.5$ is shown in FIG. \ref{f2a}(a), normalized by $-1$ to obtain a maximum value of 1. When the processes of polarization are carried out at $T_c=100$ mK, the corresponding expectation value of the number operator at equilibrium approximates null ($\bar{n}\approx 0$) for $\omega_{sc}/2\pi=6$ GHz. Obviously, there is almost no effect of thermal relaxation being observed.

The expectation value $\left\langle \sigma_{\textbf{z}}(t)\right\rangle$ for the ideal case may be fitted to an exponential function to derive an effective polarization time constant, $T_\textbf{z}$ \cite{PRL2014-112-050501}
\begin{equation}
  -\left\langle \sigma_\textbf{z}(t) \right\rangle=1-\texttt{exp}\left(-\frac{t}{T_\textbf{z}}\right)
\end{equation}
with
\begin{equation}
  T_\textbf{z}\simeq \frac{1}{\Gamma_\textbf{z}}= \frac{\kappa^2+4\Delta^2}{g^2\kappa(1+\cos\theta)^2},
\end{equation}
showing that the most efficient polarization happens when the polarization is in $\sigma_z$ direction ($\cos\theta=1$). For the case $\cos\theta<0$, we may change the matching to $\delta\omega+2\bar{\Omega}=0$, so that the polarization time is always less than $\kappa/g^2$. Effective dissipation rate in the units of $g^2/\kappa=1$ versus the dimensionless parameters $\Delta/\kappa$ and $\theta$ is shown in FIG. \ref{f2b}(b). Apparently, the effective dissipation rate increases rapidly, when the Stokes photons are on resonance with the resonator.
\begin{table}[htbp]
\begin{center}
\begin{tabular}{ccccccc}
  \hline
  $\sigma_x$ & $\texttt{Re}(\Omega)=100$& $\texttt{Im}(\Omega)=0$ & $\delta\varpi=0$& $\delta\omega= 200$ \\
  $\sigma_y$ & $\texttt{Re}(\Omega)=0$& $\texttt{Im}(\Omega)=100$ & $\delta\varpi=0$& $\delta\omega= 200$ \\
  $\sigma_z$ & $\texttt{Re}(\Omega)=0$& $\texttt{Im}(\Omega)=0$ & $\delta\varpi=200$& $\delta\omega= 200$ \\
  \hline
\end{tabular}
\end{center}
\caption{Typical energy scales (in 2$\pi$ MHz) that we consider. The polarization time of the original x,y,z-axis directions is $T_x,T_y$ and $T_z$ is about $0.8\mu$s, $0.8\mu$s and $0.2\mu$s, respectively. Here we set $g/2\pi=2$ MHz, $\kappa/2\pi=20$ MHz, $2\bar{\Omega}/2\pi=200$ MHz and $\omega_{sc}/2\pi= 6$ GHz (Validating the approximation $g\ll\kappa\ll2\bar{\Omega}$).}\label{t1}
\end{table}

In our paper, reasonable sample parameters are required to validate the Markov approximation ($\kappa \gg g$), and adhere to the two RWA¡¯s, i.e., the first one made to remove the time-dependent terms of the interaction Hamiltonian (\ref{e4}) ($\omega_c,\varpi_L,\omega_{sc} \gg g,\kappa,\Omega,\tilde{\Omega}$) and the second used to isolate the exchange term of the flux qubit and resonator of Eq. (\ref{e7-16}) ($\vert\delta\omega\vert,\vert2\bar{\Omega}\vert \gg \kappa,\vert A_{\textbf{x}}\pm iA_{\textbf{y}}\vert,\vert\Theta_\textbf{z}\vert,\vert\Theta_-\vert$). Assum that the flux qubit should be prepared in the ground states of $\sigma_x(\sigma_y,\sigma_z)$ eigenbasis, the RWA condition $\vert2\bar{\Omega}\vert \gg \vert A_{\textbf{x}}\pm iA_{\textbf{y}}\vert$ requires that
$\vert2\texttt{Re}\left[\Omega\right]\vert^2\gg\vert\frac{\delta\varpi}{2}\vert^2+\vert \texttt{Im}\left[\Omega\right]\vert^2
(\vert2\texttt{Im}\left[\Omega\right]\vert^2\gg\vert\frac{\delta\varpi}{2}\vert^2+\vert \texttt{Re}\left[\Omega\right]\vert^2,
\vert\delta\varpi\vert^2\gg\vert \texttt{Re}\left[\Omega\right]\vert^2+\vert \texttt{Im}\left[\Omega\right]\vert^2)$.
Under the experimentally reasonable parameters listed in TABLE \ref{t1}, the polarization time of the original {x,y,z}-axis directions is about $1/\Gamma_x\simeq
0.8\mu(1/\Gamma_y\simeq 0.8\mu,1/\Gamma_z\simeq 0.2\mu)$s for the ideal case ($T_c=0$), which is significantly shorter than the intrinsic energy relaxation time (and the pure dephasing time) for low-temperature flux qubit up to $20\mu$s ($10\mu$s) \cite{arXiv:1407.1346v1,arXiv:1403.3871v2}. On the other hand, the effective Rabi frequency which depends on the polarization direction, the Rabi and resonant frequencies of the microwave drive, allows a fruitfully adjustable range for experimental parameters.

\begin{figure}[b]
\centering
\includegraphics[width=1 \columnwidth]{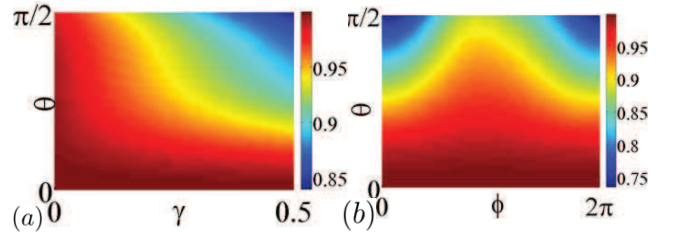}
\caption{(Color online) (a) The fidelity of the prepared state of the flux qubit, at equilibrium, versus the dimensionless parameters $\gamma=\Gamma/2\bar{\Omega}$ and $\theta$, i.e., $F_{\textbf{z}}$ vs $\gamma$ and $\theta$, for $\eta=0.52$, $\zeta=0.30$, and $\phi=\pi$; (b) $F_{\textbf{z}}$ vs $\theta$ and $\phi$, for $\eta=0.25$, $\zeta=0.17$, and $\gamma=0.02$.} \label{f3a}
\end{figure}

The purity of the generated state with an arbitrary phase factor, which is related to the Rabi frequency characteristic of the model, might be polluted by its fluctuation. To measure the reliability of the prepared state, we define the fidelity $F_\textbf{z}(t)=\left\langle - \right \vert \varrho(t) \left\vert - \right\rangle$ and plot the infidelity defined by $IF_{\textbf{z}}(t)=1-F_{\textbf{z}}(t)$ in FIG. \ref{f2c}(c) as a function of the dimensionless parameters $\Delta_R=\delta \texttt{Re}(\Omega)/\texttt{Re}(\Omega)$ and $\Delta_I=\delta \texttt{Im}(\Omega)/\texttt{Im}(\Omega)$ at equilibrium for the polarization in $\sigma_\textbf{z}=(\sigma_x+\sigma_y+\sigma_z)/\sqrt{3}$ direction with parameters $[\texttt{Re}(\Omega),\texttt{Im}(\Omega),\delta\varpi/2]/2\pi=[100,100,100]/\sqrt{3}$ MHz and $[\theta,\phi]=[\arccos(1/\sqrt{3},3\pi/4]$. It shows that, for a $20\%$ deviation of parameters $\Delta_R$ and $\Delta_I$, there is less than $1\%$ reduction in fidelity \cite{Calculation}. Thus the fidelity is slightly affected by the fluctuation of the Rabi frequency of the drive. However, we can obviously find that the polarization efficiency reduces when the unavoidable fluctuation of the Rabi frequency causes the deviation of $\left\vert \Omega \right\vert$ (i.e., $\Delta\neq 0$). As depicted in FIG. \ref{f2d}(d), for the cases $[\Delta_R,\Delta_I]=[-0.2,-0.2],[+0.2,+0.2]$, the polarization time is apparently longer than other three situations, where the dimensionless parameter $\tau=2\bar{\Omega}t$ is introduced.

\begin{figure}[t]
\centering
\includegraphics[width=1 \columnwidth]{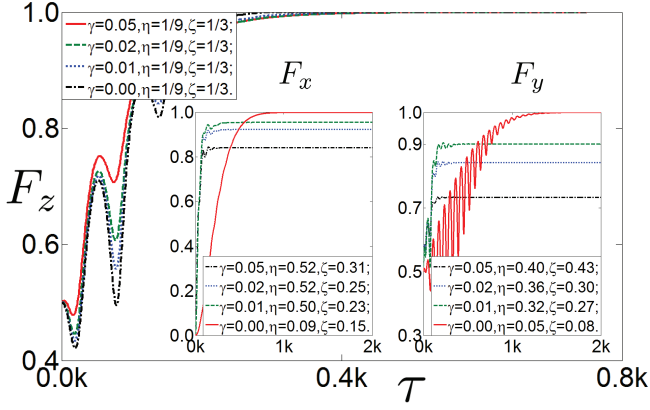}
\caption{(Color online) The evolution of the fidelity of the ground state of $\sigma_{z}$ and $\sigma_{x,y}$ (inset) for various dissipative decay rates of the flux qubit ranging from $\gamma=0$ to $\gamma=0.05$, corresponding to the enclosed optimized parameters $\eta$ and $\zeta$. Here k means $10^3$.} \label{f4}
\end{figure}

Erenow, just the resonator decay is considered. Having included the spontaneous emission and dephasing of the flux qubit, the total system and its environment can be described by the Lindblad master equation
\begin{eqnarray}
  \frac{d}{dt}\rho(t)&=&L[\tilde{H}_{1}]\rho(t)+D_{c}\rho(t)+\frac{\Gamma_s}{2}D[\sigma^{(z)}_-]\rho(t)\notag\\
      &+& \frac{\Gamma_p}{2}D[\sigma_{z}]\rho(t),
\end{eqnarray}
where $D[A]\rho=2A\rho A^{\dagger}-\{A^{\dagger}A,\rho\}$, $\Gamma_s$ is the decay rate for the spontaneous emission, and $\Gamma_p$ is the phase relaxation rate. During the numerical simulation, $\Gamma_s=\Gamma_p=\Gamma$ is assumed, and the parameters $\gamma=\Gamma/2\bar{\Omega}$, $\eta=g/2\bar{\Omega}$, and $\zeta=\kappa/2\bar{\Omega}$ are introduced. The polarization process for the flux qubit can be optimized by properly selecting the parameters $\eta$ and $\zeta$ for each combination $(\theta,\phi,\gamma)$. FIG. \ref{f3a}(a)-(b) plot the fidelity of the generated state as a function of the dimensionless  parameters (a) $\gamma=\Gamma/2\bar{\Omega}$ and $\theta$ (for $\eta=0.52$, $\zeta=0.30$, and $\phi=\pi$), and (b) $\theta$ and $\phi$ (for $\eta=0.25$, $\zeta=0.17$, and $\gamma=0.02$). The results illustrate that the fidelity can exceed the value 99\% for an optional range of the parameters. Assume the effective Rabi frequency $\bar{\Omega}=2\pi\times100$ MHz, with the choice of $\tau=500$, the polarization time in z-axis is less than $0.4\mu$s, for parameters $\eta=1/9$, $\zeta=1/3$. The evolution of the fidelity of the ground states of $\sigma_x$ and $\sigma_y$ for various dissipative decay rates of the flux qubit ranging from $\gamma=0$ to $\gamma=0.05$ is depicted in the inset of FIG. \ref{f4}. It is shown that the quality of the ground state polarization is affected by the qubit dissipation, and the case is aggravated with the increase of the intrinsic energy relaxation or pure dephasing for the qubit, especially for the state (\ref{e5}) approaching the equator on the surface of the Bloch sphere (See FIG. \ref{f3a}). In fact, according to the recent experimental data reported in \cite{arXiv:1407.1346v1,arXiv:1403.3871v2}, approximately perfect qubit polarization based upon the proposed method can be achieved. As the energy relaxation time $\frac{1}{\Gamma_s}$ and pure dephasing time $\frac{1}{\Gamma_p}$ are up to 20$\mu$s and 10$\mu$s \cite{arXiv:1407.1346v1,arXiv:1403.3871v2}, corresponding to the relatively slight $\gamma \sim \Gamma/2\bar{\Omega}\simeq 8 \times 10^{-5}$, within which the fidelity of the prepared state is almost unaffected, as shown in FIG. \ref{f3a}(a). Consequently, our scheme is in principle feasible with experimentally reasonable sample parameters.

Two main assumptions should be made in the presented theoretical model. First, we have neglected the effects of thermal relaxation of the superconducting system. No effect of thermal relaxation is observed at $T_c = 40$ mK (with $\omega_T=T_c/hk_B \simeq 0.13\times 2\pi$ GHz $\ll\omega_{sc} = 6\times 2\pi$ GHz) \cite{arXiv:1407.1346v1,arXiv:1403.3871v2}. Second, the derivation of the Markovian master equation (\ref{e9-e18}) assumes the bad resonator condition, which can be valid when the resonator dissipation rate is much larger than the coupling strength between the flux qubit and resonator in the lowest excitation manifold \cite{PRL2014-112-050501}.

In conclusion, we have demonstrated the initialization of a superconducting flux qubit assisted by a microwave resonator. The proposed technique allows any orbital state of the Bloch sphere with arbitrary phase factor of the flux qubit to be prepared by adjusting the Rabi frequency and the detunings of the drive and resonator. State preparation fidelities in excess of 99\% and the required time on the order of magnitude of microsecond are in principle possible for experimentally reasonable system parameters. Such a type of resonator-assisted qubit initialization method could find many applications in the future quantum technologies.

We are grateful to Luyan Sun and Shi-Biao Zheng for insightful discussions. This work was supported from the Major State Basic Research Development Program of China under Grant No. 2012CB921601, the National Natural Science Foundation of China under Grants No. 11405031, No. 11347114, No. 11305037, and No. 11374054, the Natural Science Foundation of Fujian Province under Grant No. 2014J05005 and No. 2013J01012, and the fund from Fuzhou University. Zhang-Qi Yin was supported by the Major State Basic Research Development Program of China under Grants No. 2011CBA00300 and No. 2011CBA00302, and the National Natural Science Foundation of China under Grants No. 11105136 and No. 11474177.

\begin{widetext}
\numberwithin{equation}{subsection}
\numberwithin{table}{subsection}
\section*{Appendix: Derivation of Markovian Master Equation and analysis of approximations}
\subsection{System Hamiltonian}
We here show all details on the derivation of Markovian master equation and analysis of approximations. Let us begin with the calculation of the interaction Hamiltonian (\ref{e8}). After moving into the interaction frame of $H_2=\bar{\Omega}\sigma_{\textbf{z}}+\delta\omega a^{\dagger}a$, the Hamiltonian (\ref{e4}) transforms to
\begin{eqnarray}
  \tilde{H}_{2}(t) &=& [ge^{itH_2}(a^{\dagger}\sigma_-)e^{-itH_2}+H.c.]+e^{itH_2}[\texttt{Re}(\Omega)\sigma_x+\texttt{Im}(\Omega)\sigma_y
           +\frac{\delta\varpi}{2} \sigma_z-\bar{\Omega}\sigma_{\textbf{z}}]e^{-itH_2} \notag \\
           &=& [e^{it\delta\omega a^{\dagger}a}a^{\dagger}e^{-it\delta\omega a^{\dagger}a}e^{it\bar{\Omega}\sigma_{\textbf{z}}}[\Theta_{\textbf{x}} \sigma_{\textbf{x}}+\Theta_{\textbf{y}} \sigma_{\textbf{y}}+\Theta_{\textbf{z}}\sigma_{\textbf{z}}]e^{-it\bar{\Omega} \sigma_{\textbf{z}}}+H.c.] \notag\\
           &+&e^{it\bar{\Omega}\sigma_{\textbf{z}}}[A_{\textbf{x}}\sigma_{\textbf{x}}+A_{\textbf{y}}\sigma_{\textbf{y}}+(A_{\textbf{z}}-\bar{\Omega}) \sigma_{\textbf{z}}]e^{-it\bar{\Omega} \sigma_{\textbf{z}}}\notag\\
           &=& [e^{i\delta\omega t}a^{\dagger}[\Theta_{\textbf{x}} e^{it\bar{\Omega}\sigma_{\textbf{z}}}\sigma_{\textbf{x}}e^{-it\bar{\Omega}\sigma_{\textbf{z}}}
           +\Theta_{\textbf{y}}e^{it\bar{\Omega} \sigma_{\textbf{z}}}\sigma_{\textbf{y}}e^{-it\bar{\Omega} \sigma_{\textbf{z}}}
           +\Theta_{\textbf{z}}\sigma_{\textbf{z}}]+H.c]\notag\\
           &+& A_{\textbf{x}}e^{it\bar{\Omega} \sigma_{\textbf{z}}}\sigma_{\textbf{x}}e^{-it\bar{\Omega} \sigma_{\textbf{z}}}+A_{\textbf{y}}e^{it\bar{\Omega} \sigma_{\textbf{z}}}\sigma_{\textbf{y}}e^{-it\bar{\Omega}\sigma_{\textbf{z}}}
           +(A_{\textbf{z}}-\bar{\Omega})\sigma_{\textbf{z}},
\end{eqnarray}
with
\begin{eqnarray}
 \Theta_{\textbf{x}}&=&\frac{1}{2}g\cos\theta e^{i\phi}, \Theta_{\textbf{y}}=-\frac{i}{2}ge^{i\phi},\Theta_{\textbf{z}}=\frac{1}{2}g\sin\theta e^{i\phi}, \\
 A_{\textbf{x}} &=&\cos\theta\cos\phi\texttt{Re}(\Omega) -\cos\theta\sin\phi\texttt{Im}(\Omega) +\frac{1}{2}\sin\theta\delta\varpi,\\
 A_{\textbf{y}} &=& \sin\phi \texttt{Re}(\Omega)+\cos\phi \texttt{Im}(\Omega),\\
 A_{\textbf{z}} &=&-\sin\theta\cos\phi\texttt{Re}(\Omega) +\sin\theta\sin\phi\texttt{Im}(\Omega) +\frac{1}{2}\cos\theta\delta\varpi,
\end{eqnarray}
i.e.,
\begin{eqnarray}
 &&[\Theta_{\textbf{x}},\Theta_{\textbf{y}},\Theta_{\textbf{z}}]^T = g\textbf{R}[1/2,-i/2,0]^T, \\
 &&[A_{\textbf{x}},A_{\textbf{y}},A_{\textbf{z}}]^T = \textbf{R}[\texttt{Re}(\Omega),\texttt{Im}(\Omega),\delta\varpi/2]^T.
\end{eqnarray}
Now we use the Baker-Campbell-Hausdorf expansion, and obtain
\begin{eqnarray}
 e^{it\bar{\Omega} \sigma_{\textbf{z}}}\sigma_{\textbf{x}}e^{-it\bar{\Omega} \sigma_{\textbf{z}}}&=& (e^{2it\bar{\Omega}}\sigma^{(\textbf{z})}_++e^{-2it\bar{\Omega}}\sigma^{(\textbf{z})}_-),\\
 e^{it\bar{\Omega} \sigma_{\textbf{z}}}\sigma_{\textbf{y}}e^{-it\bar{\Omega} \sigma_{\textbf{z}}}&=& i(e^{-2it\bar{\Omega}}\sigma^{(\textbf{z})}_--e^{2it\bar{\Omega}}\sigma^{(\textbf{z})}_+),
\end{eqnarray}
where $\sigma^{(\textbf{z})}_\pm =\frac{\sigma_\textbf{x} \pm i\sigma_\textbf{y}}{2}$ are the ladder operators in the $\textbf{z}$-basis.  Hence we obtain the interaction Hamiltonian which may be broken up in terms of frequency components
\begin{eqnarray}
   H_{I}(t)&=& \tilde{H}_0+ \tilde{H}_{\bar{\Omega}}(t)+\tilde{H}_{\textbf{z}}(t)+\tilde{H}_{-}(t)+\tilde{H}_{+}(t),\label{A10}\\
   \tilde{H}_0 &=& (A_{\textbf{z}}-\bar{\Omega}) \sigma_{\textbf{z}},\\
   \tilde{H}(t) &=& A_-e^{i2\bar{\Omega}t}\sigma_+^{(\textbf{z})}+A_+e^{-i2\bar{\Omega}t}\sigma_-^{(\textbf{z})},\\
   \tilde{H}_{\textbf{z}}(t) &=& \Theta_{\textbf{z}}e^{i\delta\omega t}a^{\dagger}\sigma_{\textbf{z}}
       +\Theta_{\textbf{z}}^*e^{-i\delta\omega t}a\sigma_{\textbf{z}},\\
   \tilde{H}_{-}(t) &=& \Theta_+e^{i\Delta_-t}a^{\dagger}\sigma_-^{(\textbf{z})}+\Theta^*_+e^{-i\Delta_- t}a\sigma_+^{(\textbf{z})},\\
   \tilde{H}_{+}(t) &=& \Theta_-e^{i\Delta_+t}a^{\dagger}\sigma_+^{(\textbf{z})}+\Theta^*_-e^{-i\Delta_+ t}a\sigma_-^{(\textbf{z})},
\end{eqnarray}
with
\begin{eqnarray}
 \Theta_\pm =\Theta_{\textbf{x}}\pm i\Theta_{\textbf{y}}, A_{\pm}=A_{\textbf{x}}\pm iA_{\textbf{y}},\Delta_{\pm} = \delta\omega \pm 2\bar{\Omega}.
 \end{eqnarray}
From here we will drop the (\textbf{z}) superscript and just note that we are working in the $\sigma_{\textbf{z}}$ eigenbasis. In order to intuitively make out the frequency components of the interaction Hamiltonian (\ref{A10}), where the parameters are listed in Table (\ref{tab}.1). So we obtain
\begin{eqnarray}\label{A17}
   H_{I}(t)&=& \sum^5_{\alpha=1}H_{\alpha}(t)=C_1e^{i\omega_1 t}A^{\dagger}_1J_{1}+\sum^5_{\alpha=2}C_\alpha e^{i\omega_\alpha t}A^{\dagger}_\alpha J_\alpha+C^*_\alpha e^{-i\omega_\alpha t}A_\alpha J^{\dagger}_\alpha.
\end{eqnarray}
\begin{table}[h]\label{tab}
\begin{center}
\begin{tabular}{cccccc}
  \hline
  $\alpha$ &1&2&3&4&5\\\hline
  $J_{\alpha}$   & $\sigma_\textbf{z}$  & $\sigma_+$  &$\sigma_\textbf{z}$  &$\sigma_-$ &  $\sigma_+$\\
  $A_\alpha$ & I & I & $a$ & $a$ & $a$\\
  $\omega_\alpha$  & 0 & $2\bar{\Omega}$ & $\delta\omega$ & $\Delta_{-}$ & $\Delta_{+}$\\
  $C_\alpha$    & $A_{\textbf{z}}-\bar{\Omega}$  & $A_{-}$ & $\Theta_{\textbf{z}}$ & $\Theta_+$ & $\Theta_-$\\
  \hline
\end{tabular}
\end{center}
\caption{Relative parameters of the system's Hamiltonian.}
\end{table}

\subsection{Derivation of Markovian master equation for the interaction Hamiltonian with multi-frequency components}
Here we use the Lindblad master equation to describe the evolution of the joint system, where the dynamics of hybrid quantum system may be depicted as an effective dissipator acting upon the flux qubit alone \cite{OUP2002}:
\begin{eqnarray}
  \dot{\rho}(t)=L\left[H_{I}(t)\right]\rho(t)+D_{c}\rho(t),
\end{eqnarray}
where $L$ is a superoperator $L[H_{I}(t)]\rho=-i[H_{I}(t),\rho]$ describing the Hermitian Hamiltonian of the system (\ref{A17}),
and $D_c$ is a dissipator describing the non-Hermitian dynamics of the system due to the coupling to a Markovian resonator
\cite{SV1974}:
\begin{eqnarray}\label{B2}
  D_{c}=\frac{\kappa}{2}\left((1+\bar{n})D[a]+\bar{n}D[a^{\dagger}]\right),
\end{eqnarray}
with $D[A]\rho=2A\rho A^{\dagger}-\{A^{\dagger}A,\rho\}$, where $\bar{n}$ is the expectation value of the photon number operator at equilibrium
\begin{equation}
\bar{n}=\frac{1}{e^{\omega_c/k_BT_c}-1},
\end{equation}
where $k_B$ is the Boltzmann constant, and $T_c$ is the temperature of the bath.

We here move to the interaction frame defined by the dissipator $D_c$. Any interaction superoperators are transformed into $\tilde{S}(t)=e^{-tD_c} S(t)e^{tD_c}$, except for the density operator $\tilde{\rho}(t)=e^{-tD_c}\rho(t)$. Then the master equation (\ref{B2}) of the hybrid quantum system is reduced to
\begin{eqnarray}
  \frac{d}{dt}\tilde{\rho}(t)=\tilde{L}[H_{I}(t)]\tilde{\rho}(t).
\end{eqnarray}

We define a projection operator $\hat{P}$ onto the relevant degrees of freedom for our reduced system
\begin{equation}
  \hat{P}\rho(t)=\varrho(t)\otimes\rho_{eq},
\end{equation}
where $\varrho(t) = \texttt{tr}_c[\rho(t)]$ is the reduced state of the flux qubit and $\rho _{eq}$ is the equilibrium state of the resonator under the dissipation, satisfying $D_c\rho _{eq}=0$. To obtain Markovian master equation for the driven flux qubit, we assume the bad resonator condition $\kappa \gg g$. Thus the reduced dynamics of the flux qubit is transformed into the second order time-convolutionless (TCL) master equation \cite{OUP2002}
\begin{equation}
  \frac{d}{dt}\hat{P}\tilde{\rho}(t)=\int_0^td\tau \hat{P}\tilde{L}[H_{I}(t)]\tilde{L}[H_{I}(t-\tau)]\hat{P}\tilde{\rho}(t).
\end{equation}

Using the following algebraic transformation of the dissipator $D^\dagger_c$, which satisfies tr$_c[D_c^{\dagger}[A]B] = $tr$_c[AD_c[B]]$ for all operators A,B on the resonator,
\begin{eqnarray}
 D_{c}^{\dagger}[\textbf{I}]&=&\textbf{0}, D_{c}^{\dagger}[a]=-\frac{\kappa}{2}a, D_{c}^{\dagger}[a^{\dagger}]=-\frac{\kappa}{2}a^{\dagger}, \\
 e^{tD_{c}^{\dagger}}[\textbf{I}]&=&\textbf{I}, e^{tD_{c}^{\dagger}}[a]=e^{-\kappa t/2}a, e^{tD_{c}^{\dagger}}[a^{\dagger}]=e^{-\kappa t/2}a^{\dagger},
\end{eqnarray}
we obtain
\begin{equation}
  \hat{P}\tilde{\rho}(t)=\texttt{tr}_c[e^{-tD_c}\rho(t)]\otimes\rho_{eq}=\texttt{tr}_c[e^{-tD^{\dagger}_c}[\textbf{I}]\rho(t)]\otimes\rho_{eq}=\hat{P}\rho(t),
\end{equation}
where we have used $D_cP\rho(t)=\texttt{tr}_c[\rho(t)]\otimes D_c\rho_{eq}=0$. Thus the reduced dynamics of the flux qubit is given by \cite{PRL2014-112-050501}:
\begin{eqnarray}\label{B10}
  \frac{d}{dt}\varrho(t)&=&\int_0^td\tau \texttt{tr}_c[L[H_{I}(t)]e^{\tau D_c}L[H_{I}(t-\tau)]\varrho(t)\otimes\rho_{eq}]\notag\\
    &=&\int_0^td\tau \texttt{tr}_c[e^{\tau D^{\dagger}_c}(L[H_{I}(t)])L[H_{I}(t-\tau)]\varrho(t)\otimes\rho_{eq}]\notag\\
    &=&-\sum_{\alpha,\delta}\int_0^td\tau A_\alpha \texttt{tr}_c[[H_{\alpha}(t),[H_{\delta}(t-\tau),
       \varrho(t)\otimes\rho_{eq}]]],
\end{eqnarray}
with
\begin{equation}
 A_\alpha=
 \begin{cases}
 1 & \alpha=1,2;\\
 e^{-\kappa\tau/2} & \alpha=3,4,5.
 \end{cases}
\end{equation}
Starting with the 2nd order TCL master equation (\ref{B10}), we now expand this in terms of the component Hamiltonians $H_{\alpha}(t)$, and define
\begin{eqnarray}
  F_{\alpha\delta}(t,t-\tau)&=&\texttt{tr}_c[[H_{\alpha}(t),[H_{\delta}(t-\tau),\varrho(t)\otimes\rho_{eq}]]].
\end{eqnarray}
Using the properties of the equilibrium state of the resonator
\begin{eqnarray}
  \texttt{tr}_c[aa^{\dagger}\rho_{eq}]&=&\bar{n}+1,\texttt{tr}_c[a^{\dagger}a\rho_{eq}]=\bar{n},\texttt{tr}_c[a^{\dagger}a^{\dagger}\rho_{eq}]=\texttt{tr}_c[aa\rho_{eq}]
  =\texttt{tr}_c[a\rho_{eq}]=\texttt{tr}_c[a^{\dagger}\rho_{eq}]=0,
\end{eqnarray}
we obtain three cases in the following. \\
(1) $\alpha,\delta=1,2$:
\begin{eqnarray}
  F_{\alpha \delta}(t,s) &=& [H_\alpha(t),[H_\delta(t),\varrho(t)]];
\end{eqnarray}
(2) $\alpha,\delta=3,4,5$:
\begin{eqnarray}
  F_{\alpha\delta}(t,s) &=& \texttt{tr}_c[C_\alpha(t)^*J^{\dagger}_\alpha A_\alpha,[C_\delta(s)J_\delta A_\delta^{\dagger},\varrho(t)\otimes\rho_{eq}]] + \texttt{tr}_c[C_\alpha(t)J_\alpha A^{\dagger}_\alpha,[C_\delta(s)^*J^{\dagger}_\delta A_\delta,\varrho(t)\otimes\rho_{eq}]] \notag\\
  &=&  \texttt{tr}_c[A^{\dagger}_\alpha A_\delta\rho_{eq}][C_\alpha(t)C_\delta(s)^*J_\alpha J^{\dagger}_\delta\varrho+C_\delta(s)C_\alpha(t)^*\varrho J_\delta J^{\dagger}_\alpha-C_\delta(s)^*C_\alpha(t)J^{\dagger}_\delta\varrho J_\alpha-C_\alpha(t)^*C_\delta(s)J^{\dagger}_\alpha\varrho J_\delta] \notag\\
  &+& \texttt{tr}_c[A_\alpha A_\delta^{\dagger}\rho_{eq}][C_\alpha(t)^*C_\delta(s)J^{\dagger}_\alpha J_\delta\varrho+C_\delta(s)^*C_\alpha(t)\varrho J^{\dagger}_\delta J_\alpha -C_\delta(s)C_\alpha(t)^*J_\delta\varrho J^{\dagger}_\alpha -C_\alpha(t)C_\delta(s)^*J_\alpha\varrho J^{\dagger}_\delta],\notag\\
  &=& \bar{n}[C_\alpha(t)C_\delta(s)^*J_\alpha J^{\dagger}_\delta\varrho+C_\delta(s)C_\alpha(t)^*\varrho J_\delta J^{\dagger}_\alpha-C_\delta(s)^*C_\alpha(t)J^{\dagger}_\delta\varrho J_\alpha-C_\alpha(t)^*C_\delta(s)J^{\dagger}_\alpha\varrho J_\delta] \notag\\
  &+&(\bar{n}+1)[C_\alpha(t)^*C_\delta(s)J^{\dagger}_\alpha J_\delta\varrho+C_\delta(s)^*C_\alpha(t)\varrho J^{\dagger}_\delta J_\alpha -C_\delta(s)C_\alpha(t)^*J_\delta\varrho J^{\dagger}_\alpha -C_\alpha(t)C_\delta(s)^*J_\alpha\varrho J^{\dagger}_\delta];
\end{eqnarray}
(3) $\alpha=1,2,\delta=3,4,5$ or $\delta=1,2,\alpha=3,4,5$:
\begin{equation}
 F_{\alpha \delta}(t,s)=0,
\end{equation}
where we suppose that the time-dependence of the Hamiltonian was included in $C_\alpha(t)=C_\alpha e^{i\omega_\alpha t}$.

To calculate the dissipator for these terms in the Markovian limit we take the upper limit of the integral to infinity $\int_0^t d\tau\rightarrow \int_0^\infty d\tau$, and define the superoperator generators
\begin{eqnarray}
  G_{\alpha,\delta}(t) \varrho(t)&=& -\int_0^\infty d\tau A_\alpha F_{\alpha \delta}(t,t-\tau), \notag\\
  G_\alpha(t) \varrho(t)&=& -\int_0^\infty d\tau A_\alpha F_{\alpha\alpha}(t,t-\tau).
\end{eqnarray}
Hence the reduced system master equation is given by
\begin{equation}\label{B18}
  \frac{d}{dt}\varrho(t)=\sum_{\alpha} G_{\alpha}(t)\varrho(t)+\sum_{\alpha\neq\delta} G_{\alpha,\delta}(t)\varrho(t),
\end{equation}
where $G_\alpha(t)$ are the diagonal terms of the master equation, while $G_{\alpha,\delta}(t)$ are the cross-terms which do not generate a completely positive map and can be removed under certain parameter regimes with an appropriate RWA.

We begin with the calculation of the diagonal terms of the master equation,
\newline
(1) $\alpha =1$
\begin{eqnarray}
G_{1}(t)\varrho(t)&=&-\int_{0}^{\infty}d\tau F_{11}(t,t-\tau)=-\int_{0}^{\infty}d\tau\lbrack\tilde{H}_{0},[\tilde{H}_{0},\varrho(t)]]\notag \\
     &=&(A_{\textbf{z}}-\bar{\Omega})^{2}\int_{0}^{\infty }d\tau \lbrack 2\sigma_{\textbf{z}}\varrho(t)\sigma_{\textbf{z}}
     -\varrho(t)\sigma_{\textbf{z}}\sigma_{\textbf{z}}-\sigma_{\textbf{z}}\sigma_{\textbf{z}}\varrho (t)]  \nonumber \\
     &=&(A_{\textbf{z}}-\bar{\Omega})^{2}\int_{0}^{\infty }d\tau D[\sigma_{\textbf{z}}]\varrho (t),
\end{eqnarray}
which won't be convergent unless the constant of the Hamiltonian is equal to zero, $i.e.$, $\bar{\Omega}=A_{\textbf{z}}$. Hence we obtain the effective Rabi frequency
\begin{equation}
  \bar{\Omega}=-\texttt{Re}\left[\Omega\right]\sin\theta\cos\phi+\texttt{Im}\left[\Omega\right]\sin\theta\sin\phi+\frac{\delta\varpi}{2}\cos\theta.
\end{equation}
\newline
(2) $\alpha =2$
\begin{eqnarray}
G_{2}(t)\varrho (t) &=&-\int_{0}^{\infty }d\tau F_{22}(t,t-\tau)=-\int_{0}^{\infty }d\tau \lbrack \tilde{H}_{\bar{\Omega}}(t),
     [\tilde{H}_{\bar{\Omega}}(t-\tau ),\varrho (t)]]  \nonumber \\
     &=&-\int_{0}^{\infty }d\tau \{[A_{-}e^{2i\bar{\Omega}t}\sigma_{+},[A_{-}e^{2i\bar{\Omega}(t-\tau )}\sigma_{+},\varrho (t)]]+[A_{-}e^{2i\bar{\Omega}t}\sigma_{+},[A_{+}e^{-2i\bar{\Omega}(t-\tau)}\sigma_{-},\varrho (t)]]  \nonumber \\
     &+&[A_{+}e^{-2i\bar{\Omega}t}\sigma_{-},[A_{-}e^{2i\bar{\Omega}(t-\tau )}\sigma_{+},\varrho (t)]]+[A_{+}e^{-2i\bar{\Omega}t}\sigma_{-},[A_{+}e^{-2i\bar{\Omega}(t-\tau)}\sigma_{-},\varrho (t)]]\}  \nonumber \\
     &=&\lambda _{2}\left\vert A_{-}\right\vert^{2}L[\sigma_{-}\sigma_{+}-\sigma_{+}\sigma_{-}]\varrho (t)-i\lambda _{2}A_{-}^{2}
     e^{4i\bar{\Omega}t}D[\sigma_{+}]\varrho (t)+i\lambda _{2}A_{+}^{2}e^{-4i\bar{\Omega}t}D[\sigma_{-}]\varrho(t),
\end{eqnarray}
with $\lambda _{2}=(2\bar{\Omega})^{-1}$. The high frequency terms $e^{\pm 4i\bar{\Omega}t}$ can be removed by making the standard RWA under the parameters regimes $4\bar{\Omega}\gg \kappa,\lambda _{2}A_{\pm}^{2}$. Hence we have
\begin{equation}
G_{2}(t)\varrho (t)=D_{0}L[\sigma_{-}\sigma_{+}-\sigma_{+}\sigma_{-}]\varrho (t),
\end{equation}
with $D_{0}=\lambda _{2}\vert A_{-}\vert ^{2}$.\newline
(3) $\alpha =3,4,5$
\begin{eqnarray}
G_{\alpha }(t)\varrho (t) &=&-\int_{0}^{\infty }d\tau e^{-\kappa \tau/2}F_{\alpha \alpha }(t,t-\tau )  \nonumber \\
    &=&-\left\vert C_{\alpha }\right\vert ^{2}\int_{0}^{\infty }d\tau e^{-\kappa\tau /2}\{(\bar{n}+1)[e^{-i\omega _{\alpha }\tau }(J_{\alpha }^{\dagger}J_{\alpha}\varrho -J_{\alpha }\varrho J_{\alpha }^{\dagger})+e^{i\omega _{\alpha }\tau}(\varrho J_{\alpha }^{\dagger}J_{\alpha }-J_{\alpha }\varrho J_{\alpha }^{\dagger})]\nonumber \\
    &+&\bar{n}[e^{i\omega _{\alpha }\tau }(J_{\alpha }J_{\alpha }^{\dagger}\varrho-J_{\alpha }^{\dagger}\varrho J_{\alpha })+e^{-i\omega _{\alpha }\tau }(\varrho J_{\alpha }J_{\alpha }^{\dagger}-J_{\alpha }^{\dagger}\varrho J_{\alpha })]\}  \nonumber\\
    &=&\left\vert C_{\alpha }\right\vert ^{2}(\bar{n}+1)[(\gamma _{\alpha}-i\lambda _{\alpha })(J_{\alpha }^{\dagger}J_{\alpha }\varrho -J_{\alpha }\varrho J_{\alpha }^{\dagger})+(\gamma _{\alpha }+i\lambda _{\alpha })(\varrho J_{\alpha}^{\dagger}J_{\alpha }-J_{\alpha }\varrho J_{\alpha }^{\dagger})]  \nonumber \\
    &+&\left\vert C_{\alpha }\right\vert ^{2}\bar{n}[(\gamma _{\alpha }+i\lambda_{\alpha })(J_{\alpha }J_{\alpha }^{\dagger}\varrho -J_{\alpha }^{\dagger}\varrho J_{\alpha })+(\gamma _{\alpha }-i\lambda _{\alpha })(\varrho J_{\alpha}J_{\alpha }^{\dagger}-J_{\alpha }^{\dagger}\varrho J_{\alpha })]  \nonumber \\
    &=&\left\vert C_{\alpha }\right\vert ^{2}\gamma _{\alpha }[(\bar{n}+1)D[J_{\alpha }]+\bar{n}D[J_{\alpha }^{\dagger}]]-\left\vert C_{\alpha}\right\vert ^{2}\lambda _{\alpha }L[(\bar{n}+1)J_{\alpha }^{\dagger}J_{\alpha }-\bar{n}J_{\alpha }J_{\alpha }^{\dagger}]\varrho (t).
\end{eqnarray}
Hence
\begin{equation}
G_{\alpha }(t)\varrho (t)=\frac{\Gamma _{\alpha }}{2}[(\bar{n}+1)D[J_{\alpha}]+\bar{n}D[J_{\alpha }^{\dagger}]]
    -\Omega _{\alpha }L[(\bar{n}+1)J_{\alpha}^{\dagger}J_{\alpha }-\bar{n}J_{\alpha }J_{\alpha }^{\dagger}],
\end{equation}
with $\Gamma _{\alpha }=2\left\vert C_{\alpha }\right\vert ^{2}\gamma_{\alpha }$ and $\Omega _{\alpha }=\left\vert C_{\alpha }\right\vert
^{2}\lambda _{\alpha }$.

The cross-terms $G_{\alpha,\delta}(t)(\alpha,\delta=3,4,5)$ will still have time dependence of $e^{\pm i(\omega_\alpha-\omega_\delta)t}$ (Other
cross-terms are all zero under the convergence condition of the master equation, i.e., $A_{\textbf{z}}=\bar{\Omega}$.). Thus, if we have $\left\vert\omega_\alpha-\omega_\delta\right\vert\gg\kappa$ for all $\alpha,\delta$, then we can make a RWA and disregard these high frequency
terms.

In this case, the master equation (\ref{B18}) reduces to:
\begin{eqnarray}
\frac{d}{dt}\varrho(t)&=& G_2(t)\varrho(t)+\sum^5_{\alpha=3}(\frac{\Gamma_{\alpha}}{2}\tilde{D}_{\alpha}
    -\Omega_{\alpha} L[\tilde{H}_{\alpha}])\varrho(t),
\end{eqnarray}
with
\begin{eqnarray}
\tilde{D}_{\alpha} &=& (\bar{n}+1)D[J_{\alpha}]+\bar{n}D[J^{\dagger}_{\alpha}],\nonumber \\
\Omega_{\alpha} &=&\frac{4\left\vert C_\alpha\right\vert^2\omega_\alpha}{\kappa^2+4\omega_\alpha^2},
\Gamma_{\alpha} = \frac{4\left\vert C_\alpha\right\vert^2\kappa}{\kappa^2+4\omega_\alpha^2},  \nonumber \\
\tilde{H}_{\alpha} &=& (\bar{n}+1)J^{\dagger}_{\alpha} J_{\alpha}-\bar{n}J_{\alpha}J^{\dagger}_{\alpha},  \nonumber
\end{eqnarray}
where $\tilde{H}_{\alpha}, \tilde{D}_{\alpha},\Omega_{\alpha},\Gamma_{\alpha}$ are the effective Hamiltonian, dissipator, frequency and
dissipation rate of model $(\alpha)$.

We consider the evolution of the flux qubit which is diagonal in the basis $\{\left\vert -\right\rangle, \left\vert +\right\rangle\}$, $\varrho (t)=\sum_{m=\pm 1}P_{m}(t)\varrho _{m}$. Here $P_{m}(t)=\left\langle m\right\vert \varrho (t)\left\vert m\right\rangle $ is the probability of finding the system in the state $\varrho_{m}=\left\vert m\right\rangle \left\langle m\right\vert $ at the arbitrary time $t$, and satisfies the equation:
\begin{equation}\label{e45}
\frac{d}{dt}P_{m}(t)=\texttt{tr}_{c}[G_{2}(t)\varrho (t)\left\vert m\right\rangle \left\langle m\right\vert ]
   +\sum_{\alpha =3}^{5}\texttt{tr}_{c}[(\frac{\Gamma_{\alpha }}{2}\tilde{D}_{\alpha }-\Omega _{\alpha }
   L[\tilde{H}_{\alpha}])\varrho (t)\left\vert m\right\rangle \left\langle m\right\vert ],
\end{equation}
with
\begin{eqnarray}
\texttt{tr}_{c}[&L&[\sigma_{-}\sigma_{+}-\sigma_{+}\sigma_{-}]\varrho (t)\left\vert m\right\rangle \left\langle m\right\vert]
        =-i\texttt{tr}_{c}[[\sigma_{-}\sigma_{+}\varrho(t)-\sigma_{+}\sigma_{-}\varrho (t)-\varrho(t)\sigma_{-}\sigma_{+}
           +\varrho(t)\sigma_{+}\sigma_{-}]\left\vert m\right\rangle \left\langle m\right\vert],\nonumber \\
        &=&-i\texttt{tr}_{c}[\varrho (t)\left\vert m\right\rangle \left\langle m\right\vert \sigma_{-}\sigma_{+}-\varrho (t)
           \left\vert m \right\rangle \left\langle m\right\vert \sigma_{+}\sigma_{-}-\varrho (t)\sigma_{-}\sigma_{+}\left\vert m\right\rangle \left\langle m\right\vert +\varrho(t)\sigma_{+}\sigma_{-}\left\vert m\right\rangle \left\langle m\right\vert ]\nonumber \\
        &=&-i\texttt{tr}_{c}[(\delta _{m,-1}-\delta _{m,1}-\delta _{m,-1}+\delta_{m,1})\varrho (t)
        \left\vert m\right\rangle \left\langle m\right\vert ]\nonumber \\
        &=&0,
\end{eqnarray}
\begin{eqnarray}
\texttt{tr}_{c}[&D&[\sigma_{+}]\varrho (t)\left\vert m\right\rangle \left\langle m\right\vert ]
        =\texttt{tr}_{c}[2\sigma_{+}\varrho (t)\sigma_{-}\left\vert m\right\rangle\left\langle m\right\vert
           -\sigma_{-}\sigma_{+}\varrho (t)\left\vert m\right\rangle \left\langle m\right\vert
           -\varrho(t)\sigma_{-}\sigma_{+}\left\vert m\right\rangle \left\langle m\right\vert ], \notag\\
        &=&\texttt{tr}_{c}[2\varrho (t)\sigma_{-}\left\vert m\right\rangle \left\langle m\right\vert \sigma_{+}
           -\varrho (t)\left\vert m\right\rangle \left\langle m\right\vert \sigma_{-}\sigma_{+}
           -\varrho (t)\sigma_{-}\sigma_{+}\left\vert m\right\rangle\left\langle m\right\vert], \notag\\
        &=&2(\delta _{m,1}-\delta _{m,-1})\texttt{tr}_{c}[\varrho (t)\left\vert -1\right\rangle \left\langle -1\right\vert ] \notag\\
        &=&2(\delta _{m,1}-\delta _{m,-1})P_{-1}(t),
\end{eqnarray}
\begin{eqnarray}
\texttt{tr}_{c}[&D&[\sigma_{-}]\varrho (t)\left\vert m\right\rangle \left\langle m\right\vert ]
        =\texttt{tr}_{c}[2\sigma_{-}\varrho (t)\sigma_{+}\left\vert m\right\rangle\left\langle m\right\vert
            -\sigma_{+}\sigma_{-}\varrho (t)\left\vert m\right\rangle \left\langle m\right\vert
            -\varrho(t)\sigma_{+}\sigma_{-}\left\vert m\right\rangle \left\langle m\right\vert ], \notag\\
        &=&\texttt{tr}_{c}[2\varrho (t)\sigma_{+}\left\vert m\right\rangle \left\langle m\right\vert \sigma_{-}
            -\varrho (t)\left\vert m\right\rangle \left\langle m\right\vert \sigma_{+}\sigma_{-}
            -\varrho (t)\sigma_{+}\sigma_{-}\left\vert m\right\rangle\left\langle m\right\vert ], \notag\\
        &=&2(\delta _{m,-1}-\delta _{m,1})P_{1}(t),
\end{eqnarray}
\begin{eqnarray}
\texttt{tr}_{c}[&D&[\sigma_{\textbf{z}}]\varrho (t)\left\vert m\right\rangle \left\langle m\right\vert ]
        =\texttt{tr}_{c}[2\sigma_{\textbf{z}}\varrho (t)\sigma_{\textbf{z}}\left\vert m\right\rangle \left\langle m\right\vert
            -\sigma_{\textbf{z}}\sigma_{\textbf{z}}\varrho (t)\left\vert m\right\rangle \left\langle m\right\vert
            -\varrho (t)\sigma_{\textbf{z}}\sigma_{\textbf{z}}\left\vert m\right\rangle \left\langle m\right\vert ], \notag\\
        &=&\texttt{tr}_{c}[2\varrho (t)\sigma_{\textbf{z}}\left\vert m\right\rangle \left\langle m\right\vert \sigma_{\textbf{z}}
            -\varrho (t)\left\vert m\right\rangle\left\langle m\right\vert \sigma_{\textbf{z}}\sigma_{\textbf{z}}
            -\varrho (t)\sigma_{\textbf{z}}\sigma_{\textbf{z}}\left\vert m\right\rangle \left\langle m\right\vert ], \notag\\
        &=&0,
\end{eqnarray}
\begin{equation}
\texttt{tr}_{c}[\tilde{D}_{3}\varrho (t)\left\vert m\right\rangle \left\langle m\right\vert ]
        =(2\bar{n}+1)D[\sigma_{\textbf{z}}]\varrho (t)\left\vert m\right\rangle \left\langle m\right\vert ]=0,
\end{equation}
\begin{eqnarray}
\texttt{tr}_{c}[&\tilde{D}_{4}&\varrho (t)\left\vert m\right\rangle \left\langle m\right\vert ]
        =(\bar{n}+1)D[\sigma_{-}]\varrho (t)\left\vert m\right\rangle \left\langle m\right\vert ]
           +\bar{n}D[\sigma_{+}]\varrho(t)\left\vert m\right\rangle \left\langle m\right\vert ] \notag\\
        &=&2(\bar{n}+1)(\delta _{m,-1}-\delta _{m,1})P_{1}(t)+2\bar{n}(\delta_{m,1}-\delta _{m,-1})P_{-1}(t),
\end{eqnarray}
\begin{eqnarray}
\texttt{tr}_{c}[&\tilde{D}_{5}&\varrho (t)\left\vert m\right\rangle \left\langle m\right\vert ]
        =(\bar{n}+1)D[\sigma_{+}]\varrho (t)\left\vert m\right\rangle \left\langle m\right\vert ]
           +\bar{n}D[\sigma_{-}]\varrho(t)\left\vert m\right\rangle \left\langle m\right\vert ] \notag\\
        &=&2(\bar{n}+1)(\delta _{m,1}-\delta _{m,-1})P_{-1}(t)+2\bar{n}(\delta_{m,-1}-\delta _{m,1})P_{1}(t),
\end{eqnarray}
\begin{eqnarray}
\texttt{tr}_{c}[&L&[\tilde{H}_{3}]\varrho (t)\left\vert m\right\rangle \left\langle m\right\vert ]
        =-i\texttt{tr}_{c}[(\bar{n}+1)\sigma_{\textbf{z}}\sigma_{\textbf{z}}\varrho(t)\left\vert m\right\rangle \left\langle m\right\vert
           -\bar{n}\sigma_{\textbf{z}}\sigma_{\textbf{z}}\varrho (t)\left\vert m\right\rangle \left\langle m\right\vert \notag\\
        &-&(\bar{n}+1)\varrho (t)\sigma_{\textbf{z}}\sigma_{\textbf{z}}\left\vert m\right\rangle \left\langle m\right\vert
           +\bar{n}\varrho (t)\sigma_{\textbf{z}}\sigma_{\textbf{z}}\left\vert m\right\rangle \left\langle m\right\vert ] \notag\\
        &=&-i\texttt{tr}_{c}[(\bar{n}+1)\varrho (t)\left\vert m\right\rangle \left\langle m\right\vert \sigma_{\textbf{z}}\sigma_{\textbf{z}}
           -\bar{n}\varrho (t)\left\vert m\right\rangle \left\langle m\right\vert \sigma_{\textbf{z}}\sigma_{\textbf{z}}\notag\\
        &-&(\bar{n}+1)\varrho (t)\sigma_{\textbf{z}}\sigma_{\textbf{z}}\left\vert m\right\rangle\left\langle m\right\vert
           +\bar{n}\varrho (t)\sigma_{\textbf{z}}\sigma_{\textbf{z}}\left\vert m\right\rangle \left\langle m\right\vert ] \notag\\
        &=&0,
\end{eqnarray}%
\begin{eqnarray}
\texttt{tr}_{c}[&L&[\tilde{H}_{4}]\varrho (t)\left\vert m\right\rangle \left\langle m\right\vert]
        =-i\texttt{tr}_{c}[(\bar{n}+1)\sigma_{+}\sigma_{-}\varrho (t)\left\vert m\right\rangle \left\langle m\right\vert
           -\bar{n}\sigma_{-}\sigma_{+}\varrho(t)\left\vert m\right\rangle \left\langle m\right\vert \notag\\
        &-&(\bar{n}+1)\varrho (t)\sigma_{+}\sigma_{-}\left\vert m\right\rangle \left\langle m\right\vert
           +\bar{n}\varrho (t)\sigma_{-}\sigma_{+}\left\vert m\right\rangle\left\langle m\right\vert ] \notag\\
        &=&-i\texttt{tr}_{c}[(\bar{n}+1)\varrho (t)\left\vert m\right\rangle \left\langle m\right\vert \sigma_{+}\sigma_{-}
           -\bar{n}\varrho (t)\left\vert m\right\rangle\left\langle m\right\vert \sigma_{-}\sigma_{+}\notag\\
        &-&(\bar{n}+1)\varrho(t)\sigma_{+}\sigma_{-}\left\vert m\right\rangle \left\langle m\right\vert
           +\bar{n}\varrho (t)\sigma_{-}\sigma_{+}\left\vert m\right\rangle \left\langle m\right\vert] \notag\\
        &=&-i\texttt{tr}_{c}[\delta _{m,1}(\bar{n}+1)\varrho (t)\left\vert 1\right\rangle\left\langle 1\right\vert
           -\delta_{m,-1}\bar{n}\varrho (t)\left\vert -1\right\rangle \left\langle -1\right\vert \notag\\
        &-&\delta _{m,1}(\bar{n}+1)\varrho (t)\left\vert 1\right\rangle \left\langle 1\right\vert
           +\delta _{m,-1}\bar{n}\varrho (t)\left\vert -1\right\rangle \left\langle -1\right\vert ] \notag\\
        &=&0,
\end{eqnarray}
\begin{eqnarray}
\texttt{tr}_{c}[&L&[\tilde{H}_{5}]\varrho (t)\left\vert m\right\rangle \left\langle m\right\vert ]
        =-i\texttt{tr}_{c}[(\bar{n}+1)\sigma_{-}\sigma_{+}\varrho (t)\left\vert m\right\rangle \left\langle m\right\vert
           -\bar{n}\sigma_{+}\sigma_{-}\varrho(t)\left\vert m\right\rangle \left\langle m\right\vert \notag\\
        &-&(\bar{n}+1)\varrho (t)\sigma_{-}\sigma_{+}\left\vert m\right\rangle \left\langle m\right\vert
           +\bar{n}\varrho (t)\sigma_{+}\sigma_{-}\left\vert m\right\rangle\left\langle m\right\vert ] \notag\\
        &=&-i\texttt{tr}_{c}[(\bar{n}+1)\varrho (t)\left\vert m\right\rangle \left\langle m\right\vert \sigma_{-}\sigma_{+}
           -\bar{n}\varrho (t)\left\vert m\right\rangle\left\langle m\right\vert \sigma_{+}\sigma_{-}\notag\\
        &-&(\bar{n}+1)\varrho(t)\sigma_{-}\sigma_{+}\left\vert m\right\rangle \left\langle m\right\vert
           +\bar{n}\varrho (t)\sigma_{+}\sigma_{-}\left\vert m\right\rangle \left\langle m\right\vert] \notag\\
        &=&-i\texttt{tr}_{c}[\delta _{m,-1}(\bar{n}+1)\varrho (t)\left\vert -1\right\rangle\left\langle -1\right\vert
           -\delta_{m,1}\bar{n}\varrho (t)\left\vert 1\right\rangle \left\langle 1\right\vert \notag\\
        &-&\delta _{m,-1}(\bar{n}+1)\varrho (t)\left\vert -1\right\rangle \left\langle -1\right\vert
           +\delta _{m,1}\bar{n}\varrho (t)\sigma_{+}\sigma_{-}\left\vert 1\right\rangle\left\langle 1\right\vert ] \notag\\
        &=&0.
\end{eqnarray}
Defining $\vec{P}(t)=(P_{-1}(t),P_{1}(t))$, the master equation (\ref{e45}) reduces to a rate equation for the state populations:
\begin{equation}
\frac{d}{dt}\vec{P}(t)=\sum_{\alpha =4,5}\Gamma _{\alpha }M_{J}^{\alpha }\vec{P}(t),
\end{equation}%
with
\begin{equation}
M_{J}^{4}\mathtt{=}\left[
\begin{array}{cc}
-\bar{n} & \bar{n}+1 \\
\bar{n} & -(\bar{n}+1)
\end{array}
\right] ,M_{J}^{5}\mathtt{=}\left[
\begin{array}{cc}
-(\bar{n}+1) & \bar{n} \\
\bar{n}+1 & -\bar{n}
\end{array}
\right].
\end{equation}
\end{widetext}


\begin{thebibliography}{99}
\bibitem{RMP2013-85-553} H. Ritsch, P. Domokos, F. Brennecke, and T. Esslinger, Rev. Mod. Phys. \textbf{85}, 553 (2013).
\bibitem{RMP2013-85-623} Z.-L. Xiang, S. Ashhab, J. Q. You, and F. Nori, Rev. Mod. Phys. \textbf{85}, 623 (2013).

\bibitem{S2011-332-1059} P. Schindler, J. T. Barreiro, T. Monz, V. Nebendahl, D. Nigg, M. Chwalla, M. Hennrich1, and R. Blatt, Science \textbf{332}, 1059 (2011).
\bibitem{N2012-482-382} M. D. Reed, L. DiCarlo, S. E. Nigg,  L. Sun, L. Frunzio, S. M. Girvin, and R. J. Schoelkopf et al., Nature \textbf{482}, 382 (2012).



\bibitem{PRL2010-105-140501} D. I. Schuster, A. P. Sears, E. Ginossar, L. DiCarlo, L. Frunzio, J. J. L. Morton, H. Wu, G. A. D. Briggs, B. B. Buckley, D. D. Awschalom, and R. J. Schoelkopf, Phys. Rev. Lett. \textbf{105}, 140501 (2010).
\bibitem{PRL2010-105-140502} Y. Kubo, F. R. Ong, P. Bertet, D. Vion, V. Jacques, D. Zheng, A. Dreau, J. F. Roch, A. Auffeves, F. Jelezko, J. Wrachtrup, M. F. Barthe, P. Bergonzo, and D. Esteve, Phys. Rev. Lett. \textbf{105}, 140502 (2010).


\bibitem{OUP1961}  A. Abragam, \emph{The Principles of Nuclear Magnetism} (Oxford University Press, New York, 1961).


\bibitem{NP2008-4-612} M. Grajcar, S. H. W. van der Ploeg, A. Izmalkov, E. Il'ichev, H.-G. Meyer, A. Fedorov, A. Shnirman, and G. Sch\"{o}n, Nat. Phys. \textbf{4}, 612 (2008).
\bibitem{PRL2003-90-133602} J. McKeever, J. R. Buck, A. D. Boozer, A. Kuzmich, H.-C. N\"{a}gerl, D. M. Stamper-Kurn, and H. J. Kimble, Phys. Rev. Lett. 90, 133602 (2003).
\bibitem{N2011-480-500} W. S. Bakr, P. M. Preiss, M. E. Tai, R. Ma, J. Simon, and M. Greiner, Nature \textbf{480}, 500 (2011).
\bibitem{PRL2008-100-140501} C. A. Ryan, O. Moussa, J. Baugh, and R. Laflamme, Phys. Rev. Lett. \textbf{100}, 140501 (2008).
\bibitem{NP2005-1-122} S. Nu{\ss}mann, K. Murr, M. Hijlkema, B. Weber, A. Kuhn, and G. Rempe, Nat. Phys. \textbf{1}, 122 (2005).
\bibitem{PRL2014-112-050501} C. J. Wood, T. W. Borneman, and D. G. Cory, Phys. Rev. Lett. \textbf{112}, 050501 (2014).
\bibitem{NL2004-428-50} P. Maunz, T. Puppe, I. Schuster, N. Syassen, P. W. H. Pinkse, and G. Rempe, Nature (London) \textbf{428}, 50 (2004).

\bibitem{PRL2009-103-103001}  D. R. Leibrandt, J. Labaziewicz, V. Vuletic, and I. L. Chuang, Phys. Rev. Lett. \textbf{103}, 103001 (2009).

\bibitem{NL2006-444-71} O. Arcizet, P.-F. Cohadon, T. Briant, M. Pinard, and A. Heidmann, Nature (London) \textbf{444}, 71 (2006).
\bibitem{NL2006-443-193} A. Naik, O. Buu, M. D. LaHaye, A. D. Armour, A. A. Clerk, M. P. Blencowe, and K. C. Schwab, Nature (London) \textbf{443}, 193 (2006).
\bibitem{PRA2010-82-041804} N. Brahms and D. M. Stamper-Kurn, Phys. Rev. A \textbf{82}, 041804 (2010).

\bibitem{PRL2012-109-183602} K. W. Murch, U. Vool, D. Zhou, S. J. Weber, S. M. Girvin, and I. Siddiqi, Phys. Rev. Lett. \textbf{109}, 183602 (2012).

\bibitem{arXiv:1407.1346v1} J.-L. Orgiazzi, C. Deng, D. Layden, R. Marchildon, F. Kitapli, F. Shen, M. Bal, F. R. Ong, and A. Lupascu, arXiv:1407.1346v1.
\bibitem{arXiv:1403.3871v2} M. Stern, G. Catelani, Y. Kubo, C. Grezes, A. Bienfait, D. Vion, D. Esteve, and P. Bertet, arXiv:1403.3871v2.
\bibitem{S1999-285-1036} J. E. Mooij, T. P. Orlando, L. Levitov, L. Tian, Caspar H. van der Wal, and S. Lloyd, Science \textbf{285}, 1036 (1999).

\bibitem{NP2010-6-772} T. Niemczyk, F. Deppe, H. Huebl, E. P. Menzel, F. Hocke, M. J. Schwarz,  J. J. Garcia-Ripoll, D. Zueco, T. H¨¹mmer, E. Solano, A. Marx, and R. Gross, Nat. Phys. \textbf{6}, 772 (2010).

\bibitem{PRL2010-105-023601} B. Peropadre, P. Forn-D\'{\i}az, E. Solano, and J. J Garc\'{\i}a-Ripoll, Phys. Rev. Lett.  \textbf{105}, 023601 (2010).



\bibitem{OUP2002} H.-P. Breuer and F. Petruccione, \emph{The Theory of Open Quantum Systems} (Oxford University Press, New York, 2002).
\bibitem{SV1974} G. S. Agarwal, \emph{Quantum Optics} (Springer-Verlag, Berlin, 1974).
\bibitem{Calculation} We have calculated the ideal ground states for the system parameters $[\texttt{Re}(\Omega),\texttt{Im}(\Omega),\delta\varpi/2]/2\pi$ with $[80,120,100]$MHz, $[120,80,100]$MHz, $[80,80,100]$MHz, and $[120,120,100]$MHz, respectively. The obtained fidelities are only reduced to 99.13\%, 99.13\%, 98.83\%, and 99.31\%, respectively, for a $20\%$ deviation of parameters $\Delta_R$ and $\Delta_I$, also reliable enough as the case with the ground state of $\sigma_\textbf{z}=(\sigma_x+\sigma_y+\sigma_z)/\sqrt{3}$.
\end{thebibliography}
\end{document}